\documentstyle[11pt,epsf]{article}
\begin{document}
\title{DYNAMIC CONDUCTANCE IN QUANTUM HALL SYSTEMS}
\author{M. B\"{U}TTIKER AND T. CHRISTEN \\
Universit\'e de Gen\`eve, 24 Quai E. Ansermet, CH-1211 Geneva\\ 
Switzerland}
\maketitle
\begin{abstract}
In the framework of the edge-channel picture and the scattering
approach to conduction, we discuss the low frequency
admittance of quantized Hall samples up to second order in frequency.
The first-order term gives the leading order phase-shift
between current and voltage and is associated with the displacement
current. It is determined by the emittance which is a capacitance in
a capacitive arrangement of edge channels but which is inductive-like
if edge channels predominate which transmit charge between {\it different}
reservoirs.
The second-order term is associated with the charge relaxation. 
We apply our results to a Corbino disc and to two- and four-terminal
quantum Hall bars, and we discuss the symmetry properties of the current response.
In particular, we calculate the longitudinal resistance and the Hall resistance
as a function of frequency.
\end{abstract}
\section{Introduction}
\label{I}
In this work we discuss the low frequency response of two-dimensional conductors
subject to a perpendicular quantizing magnetic field\cite{Klit1} 
and subject to time-dependent oscillating voltages applied to the contacts
of the sample. We use an approach\cite{CB1}$^{\!-\,}$\cite{BTP1} closely
related to the scattering approach to dc conduction\cite{MBR} in a
mesoscopic phase-coherent multiterminal sample. The approach treats current and
voltage contacts on equal footing and views the entire sample and
the nearby conductors as one electric entity.\\ \indent
It is well-known from mesoscopic conduction theory
that all dc resistances are rational
functions of transmission probabilities for carriers to reach a contact
if they are  injected at another contact.\cite{MBR} Moreover, only the states at the
Fermi surface which connect different contacts count.
In a conductor subject to a quantizing magnetic field the only states
at a Hall plateau which provide such a connection are the {\it edge states}
which are the quantum mechanical generalization of the semi-classical
`skipping orbits' of the charge carriers along the sample boundary.
As a consequence the dc conductance is quantized and determined by the 
topology of the edge-state arrangement.\cite{MBR} For a lucid discussion of the
current patterns and Hall potentials in the presence of transport, we refer the reader
to a recent work by Komiyama and Hirai.\cite{Komi} \\ \indent
Most of the early work on edge states relies on
a picture valid for noninteracting electrons which must
be modified in very pure samples with smooth
boundaries.\cite{Chkl1} Interaction leads
to a decomposition of the electron gas into compressible and incompressible
regions, where the former ones are associated with the edge states.
Interaction must also be considered to describe edge states in the
fractional quantum-Hall regime.\cite{Been1} 
A detailed description of the contacts
for chiral edge-state models has recently been provided by 
Kane and Fisher.\cite{Kane1} While in these works 
the main concern is with short-range interactions, we emphasize
in our work the long-range part of the Coulomb interaction. For
this case, edge states are important since they
contribute to screening, in contrast to the bulk
states.\cite{Komi,McDo2} Due to the Coulomb interaction,
the ac response is not only determined
by the topology of the edge-state arrangement but also by its
geometry.\\ \indent
In the following, we concentrate on a two-dimensional electron gas
with spin-split Landau levels. Coulomb interaction is taken
into account by consideration of
the capacitances of all edge channels and all gates. We
assume that additional contributions due to parasitic capacitances
associated with contacts can be neglected, that the contacts
are quantized, and that inter-edge scattering is absent.
\section{Low-frequency admittance}
\label{LFA}
\subsection{Current conserving and gauge-invariant electric transport}
\label{CCGI}
The gates together with the edge channels which are separated
by regions of incompressible electron gas can be seen as an arrangement
of mesoscopic conductors embedded in a dielectric medium.
We describe an $N$-terminal sample at a Hall plateau by
an arrangement of such conductors $j= 1, ..., M$
connected to contacts $\alpha = 1, ...,N$
(see, e.g., Fig. \ref{fig1}). We are interested in the dynamical conductance
$G_{\alpha\beta} (\omega) $ which determines the Fourier amplitudes
of the current, $\delta I_{\alpha}$,
at a contact $\alpha$ in response to an oscillating voltage 
$\delta V_{\beta} \exp(-i\omega t)$ at contact $\beta$
\begin{equation}
\delta I_{\alpha} = \sum _{\beta}
G_{\alpha \beta} (\omega) \; \delta V_{\beta}\;\;.
\label{eq1}   
\end{equation} 
The voltage variation $\delta V_{\beta}$ is related to the variation
of the electro-chemical
potential $\delta \mu _{\beta} $ in reservoir $\beta $ by  $\delta
\mu _{\beta}=e\delta V _{\beta}$, where $e$ is the electron
charge.\\ \indent
Since all the present gates and conductors are
included in the description, and due to the long-range nature of
the Coulomb interaction the theory for the conductance has to satisfy
two basic requirements, namely {\em current conservation} and
{\em gauge invariance}. Current
conservation corresponds to zero total current, $\sum _{\alpha}
\delta I_{\alpha}=0$. Hence the total charge is conserved. Gauge invariance
means that a global voltage shift, $\delta V_{\alpha} \mapsto 
\delta V_{\alpha} + \delta V$, does not affect the currents.
In particular, we fix the voltage scale such that 
$\delta V_{\alpha} \equiv 0$ corresponds to the equilibrium state
where all electrochemical potentials $\mu _{\alpha}$ are equal.
Current conservation and gauge invariance demand the admittance
matrix $G_{\alpha \beta}$ to satisfy
$\sum_{\alpha}G_{\alpha \beta}=\sum_{\beta}G_{\alpha \beta}=0$.
For example, the admittance of a two-terminal device is determined by a single
scalar quantity, $G_{\alpha \beta}= (-1)^{\alpha +\beta}G$.
Due to microreversibility the linear response coefficients
must additionally satisfy
the Onsager-Casimir reciprocity relations $G_{\alpha \beta}|_{B}
=G_{\beta \alpha}|_{-B}$ where $B$ is the magnetic field.
We are interested in the low-frequency limit and write
\begin{equation}
G_{\alpha \beta} (\omega)= G_{\alpha \beta} ^{(0)}-i\omega E_{\alpha \beta}
+ \omega^{2} K_{\alpha \beta}  + {\cal O}(\omega ^{3}) \;\; .
\label{eq2} 
\end{equation}
While the theory for the dc-conductances $G_{\alpha \beta }^{(0)}$
of quantized Hall systems is well established,\cite{MBR}
the \em emittance \rm matrix $E_{\alpha \beta}
\equiv i(dG_{\alpha \beta}/d\omega)_{\omega = 0}$
has been investigated only recently.\cite{CB1}
Below we recall these results and, in addition, provide
new results for the second order term $K_{\alpha \beta}\equiv 
(1/2)\:(d^{2}G_{\alpha \beta}/d\omega ^{2})_{\omega =0}$. 
To get some feeling for the emittance and the second order term,
consider a two-terminal device with a macroscopic capacitor $C$ in series
with a resistor $R$. For this purely capacitive structure with zero
dc-conductance one finds that the emittance is the
capacitance, $E = C$, and that  $K=RC^{2}$ contains the
$RC$ time constant associated with charge relaxation.
These simple results must be modified for {\em mesoscopic}
conductors and conductors which
connect {\em different} reservoirs.\cite{BJPC} Firstly, it is not
the geometrical capacitance but rather the {\em electro-chemical
capacitance} which relates charges on mesoscopic
conductors to voltage variations
in the reservoirs. Secondly, conductors which connect
different reservoirs allow a transmission of charge which leads to
inductance-like contributions to the emittance.
\subsection{Transmission properties of edge channels}
\label{TPOEC}
An important property of a single edge channel is
its uni-directional transparency and the absence of
backscattering.\cite{MBR} The scattering matrix $s_{\alpha \beta}$
relates the out-going current amplitude at contact $\alpha $
to the incoming current amplitude at contact $\beta $. 
The part of the scattering matrix
associated with edge channel $j$ can be written in the form
$s_{\alpha \beta}^{(j)}= \Delta
_{\alpha j}t_{j} \Delta _{j \beta}$ with a scattering amplitude
$t_{j}=\exp(i \phi _{j} (E))$. The energy dependence of the
phase $\phi _{j} (E)$ determines the
density of states of the edge channel, $dN_{j}/dE=
(1/2\pi)(d\phi _{j}/dE)$ at the Fermi energy. We mention that
for noninteracting particles the density of states is related to
the electric field at the edge.\cite{MBR}
$\Delta _{j \alpha}$ and $\Delta _{\alpha j}$ are the injection and
emission probability, respectively, of channel $j$ at contact $\alpha $:
\begin{eqnarray}
\Delta _{j\alpha} & = & \cases{1 &${\rm if \; contact } \;\; \alpha
{\rm \;\; injects \; into \; channel \;\; } j$ \cr 
                          0 &$\;\;\;\; {\rm otherwise} $ \cr} 
\label{eq3}  \\
\Delta _{\alpha j} & =& \cases{1 &${\rm if \; channel} \;\; j
{\rm \;\; emits \; into \; contact \;\; } \alpha $ \cr 
                          0 &$\;\;\;\; {\rm otherwise} $ \cr} \;\; .
\label{eq4}
\end{eqnarray}
Since an edge channel is always connected to one contact at each of its ends,
one has $\sum _{\alpha}\Delta _{k\alpha} = \sum _{\alpha}\Delta _{\alpha k} =1$.
Furthermore, $\sum _{k}\Delta _{k\alpha} $ ($=\sum _{k}\Delta _{\alpha k} $) is the
number of edge channels which enter (leave) the sample at contact $\alpha $.  
Clearly, the transmission probabilities are functions of the magnetic
field $B$ and obey the micro-reversibility conditions
$\Delta _{\alpha k}(B)= \Delta _{k \alpha}(-B)$. This follows directly
from the fact that an inversion of the magnetic field
inverses the arrows attached to the edge states. In terms of
the $\Delta _{j \alpha}$ and $\Delta _{\alpha j}$ the dc conductance
becomes\cite{CB1}
\begin{equation}
G_{\alpha \beta}^{(0)}=
\frac{e^{2}}{h} \sum _{k} \biggl( \Delta _{k\alpha}\: \Delta _{k\beta}
- \Delta _{\alpha k} \Delta _{k\beta} \biggr)\;\;.
\label{eq5} 
\end{equation}
Hence the topology of the edge-channel arrangement determines the
dc conductance. Note that current conservation, gauge invariance,
and the Onsager-Casimir relations follow immediately from the
properties of $\Delta _{j \alpha}$ and $\Delta _{\alpha j}$. 
\subsection{Screening properties: a discrete model}
\label{SPADM}
While the dc-conductance is completely determined by the overall
transmission, the ac-conductance contains information on the
nonequilibrium state within the sample. Application of an oscillatory 
voltage at a sample contact leads to a periodic charging
of the sample, and the induced electric fields 
have to be taken into account self-consistently.
In order to treat the dynamic charge and potential distribution
in the nonequilibrium state we spatially discretize the
two-dimensional electron gas. More concrete, we assign to each
edge channel $j$ a single electrochemical potential of the carriers, a single
electrical potential, and a total charge. We neglect spatio-temporal
effects inside edge channels such as edge plasmons with finite wave
numbers.\cite{Taly,Som1} However, our discussion takes into account the zero
wave-number modes of the edge states.  The electrochemical potential shift
$\delta V_{j}$ in edge channel $j$ can be written in terms of the
voltages $ \delta V_{\beta}$ in the reservoirs, $\delta V _{j}=\sum_{\beta}\Delta
_{j\beta}\: \delta V_{\beta}$. To simplify notation we write
quantities associated with edge channels as
$M$-dimensional vectors, e.g. ${\bf \delta V}=(\delta V_{1},...,
\delta V_{M})^{t}$. The nonequilibrium distribution of charges
and electric potentials
${\bf \delta q}=(\delta q_{1},...,\delta q_{M})^{t}$
and ${\bf \delta U}=(\delta U_{1},...,\delta U_{M})^{t}$, respectively,
are related by ${\bf \delta q}={\bf C \: \delta U}$, where ${\bf C}$
is the geometrical capacitance matrix 
for the edge channels and gates. As usual, ${\bf C}$
can be obtained from the Poisson equation. A relation between
the charges and the voltages of the reservoirs, ${\bf \delta q}= {\bf
C_{\mu}\; \delta  V}$, defines
the electrochemical capacitance matrix ${\bf C_{\mu} }$. Note that while
${\bf C}$ is real and frequency independent,
${\bf C_{\mu} }(\omega )$ is a function of frequency.
Here, we define ${\bf C}$ by assuming that each edge channel
is connected to its own reservoir, hence charge conservation and gauge
invariance
imply for both ${\bf C}$ and ${\bf C}_{\mu}(\omega )$ the sum rules 
$\sum _{k}C_{kl}=\sum _{l}C_{kl}=\sum _{k}C_{\mu ,kl}=\sum _{l}C_{\mu ,kl}=0$.
Both matrices are symmetric and even functions of the magnetic
field. Further we need the dynamic densities of states of the
edge channels. In the present notation they can be summarized in a
diagonal matrix ${\bf D}(\omega)$ with elements
$D_{jj}(\omega)=e^{2}(dN_{j}/dE)(1+i\tau _{j}\omega/2) +{\cal O}(\omega ^{2})$.\cite{Pret}
Here, $\tau _{j}=h(dN_{j}/dE)$ is the dwell time of carriers
in edge channel $j$. The specific form of
the $D_{jj}(\omega)$ can be understood by noticing that the charging
of edge channel $j$ is a relaxation process with a time $\tau _{j}/2$.
The nonequilibrium charge is
${\bf \delta q}= {\bf D}({\bf \delta V} -{\bf \delta U})
={\bf C \delta U}= {\bf C}_{\mu}{\bf \delta V}$ from which one
obtains the dynamic electrochemical capacitance
\begin{equation}
{\bf C }_{\mu}(\omega) ={\bf C}_{\mu}(0)+i\omega \frac{h}{2e^{2}}{\bf C}
_{\mu}^{2}(0) +{\cal O}(\omega ^{2})\;\; ,
\label{eq6}
\end{equation}
where ${\bf C}_{\mu}(0)=[{\bf 1}+{\bf CD}^{-1}(0)]^{-1}{\bf C}$
is the static part of the electrochemical capacitance matrix.
The dependence of the effective capacitance ${\bf C}_{\mu}(0)$
on the densities of states is reasonable: small densities of
states make it more difficult to charge the conductor.
For large densities of states, on the other hand, one recovers the
geometrical capacitance.
\subsection{The dynamic response}
\label{DR}
The low frequency admittance can be calculated with the help of
a recent theory which generalizes the scattering approach to
ac conduction.\cite{BTP1} The ingredients which we need are the current
response to an external voltage shift $\delta {\bf V}$ and
the current response to an internal voltage shift $\delta {\bf U}$
for carriers at the Fermi energy $E_{F}$ and for carriers
with energies
$E_{F}+\hbar \omega $. We do not enter into details here but
present only the result. We find
\begin{equation}
G_{\alpha\beta } (\omega) = G_{\alpha \beta}^{(0)}
 -  i \omega 
\sum_{k,l=1}^{N} \Delta _{\alpha k} \:
C_{\mu ,kl}(\omega)\: \Delta _{l\beta}\;\;.
\label{eq7}
\end{equation}
Using the micro-reversibility properties of the $\Delta _{\alpha k}$
and the symmetry of the capacitance matrix
it is again easy to prove current conservation, gauge invariance,  
and the Onsager-Casimir reciprocity relations. The results
(\ref{eq6}) and (\ref{eq7}) yield the admittance to order ${\cal O}(\omega
^{2})$. The emittance $E_{\alpha \beta}= \sum _{kl}\Delta _{\alpha
k}$ $C_{\mu, kl}(0)\Delta _{l \beta} $ can be understood as the sum
of all `two-body' Coulomb interactions between edge
states and/or gates.\cite{CB1} In a pure capacitive
arrangement where $\Delta _{k \alpha} = \Delta _{\alpha k}$ the emittance
matrix is a capacitance matrix, i.e. diagonal elements are positive
and off-diagonal elements are negative. As we will see below,
this can change in the presence of edge channels which connect different
reservoirs. The result for the second-order term,
$K_{\alpha \beta}= (h/2e^{2}) $ $\sum _{kjl}\Delta _{\alpha
k}C_{\mu, kj}(0)$ $ C_{\mu, jl}(0)\Delta _{l \beta} $, can thus similarly
be interpreted as a `three-body' interaction with an intermediate
step via edge channel (or gate) $j$. This simple result,
of course, relies on the discrete model and
is modified as soon as one includes internal spatio-temporal
edge excitations.
%
\begin{figure}[t]
\hspace{10mm}
 \epsfxsize = 50 mm
 \epsffile{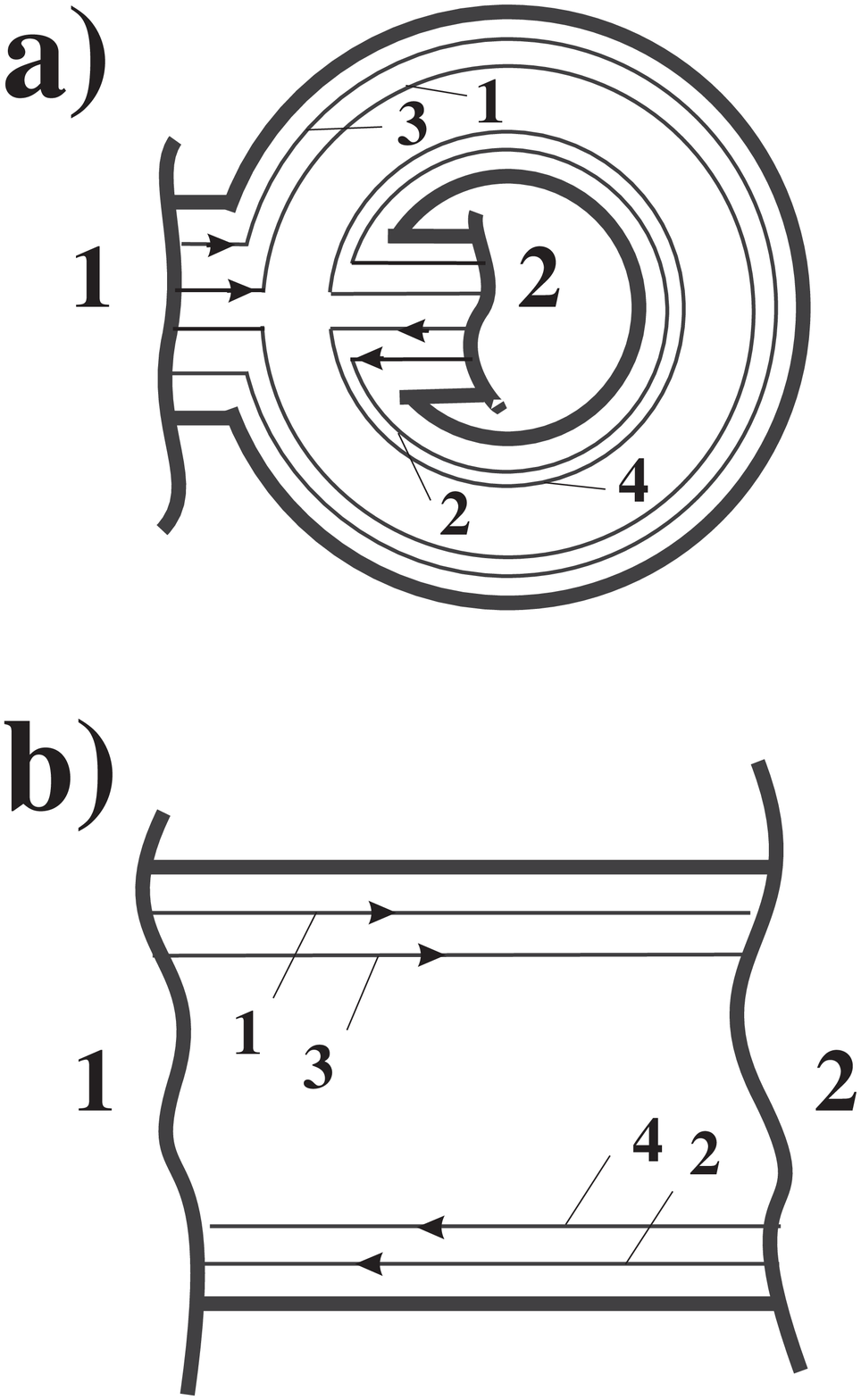}
 \epsfxsize = 50 mm
 \epsffile{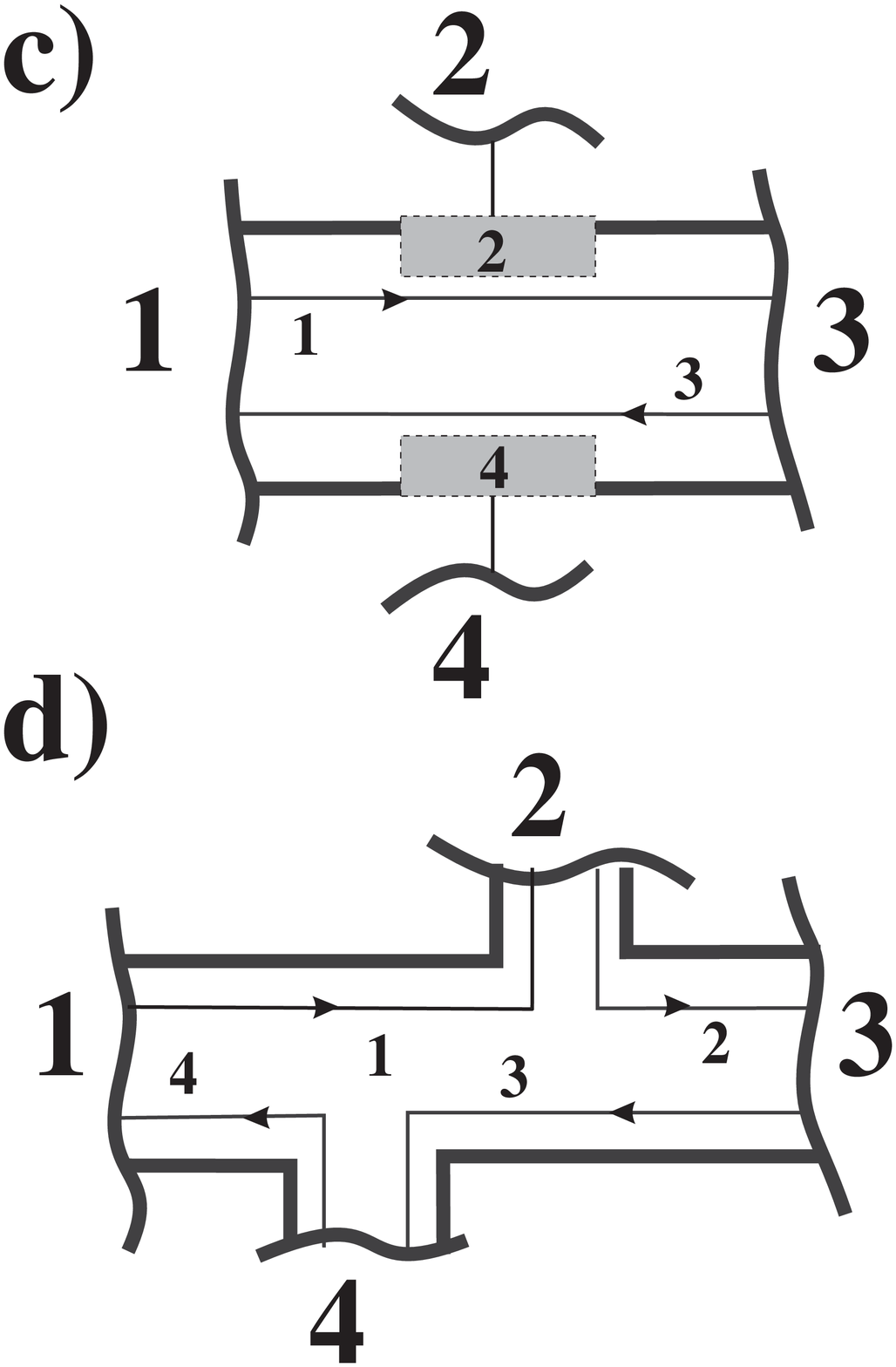}
\caption{ Two-terminal samples:
a) Corbino disc and b) quantum Hall bar. Four-terminal samples:
c) two-terminal Hall bar with two additional
gates; d) four-terminal Hall bar}
\label{fig1}
\end{figure}
%
%
%
\section{Examples}
\label{Examples}
\subsection{Two-terminal devices}
\label{Two-terminal devices}
Let us discuss the Corbino disc and the quantum Hall bar with $m$ channels 
along an edge as  shown in Fig. \ref{fig1} for a magnetic field $B$ such that $m=2$.
Channels at the upper
and the lower edges are labeled by odd and even numbers, respectively. Equation (\ref{eq6})
permits now to treat the case where different edge states are at different potentials.
For the sake of clarity, however, we assume
that the capacitances between edge channels at the same edge are very large
such that they are at equal potentials. We take thus each set together to form a single
$m$-channel conductor associated with only even or only odd labels.
The geometric capacitance between the two sets of channels is
denoted by $C$ ($=\sum _{kl}C_{2k\: 2l}$). While for the disc the reservoirs are capacitively
coupled, for the bar all edge channels transmit charge between different reservoirs.
Hence, the dc-conductance of the disc and the bar are $G(0)=0$ and
$G(0)=me^{2}/h$, respectively. The dynamic capacitance between the two sets
of edge channels is $C_{\mu}(\omega)=  C_{\mu}(0) +i\omega R C_{\mu}^{2}$
with a static electrochemical capacitance given by
$1/C_{\mu}(0)=1/C+(\sum _{{\rm even}}D_{i})^{-1} +
(\sum _{{\rm odd}}D_{i})^{-1}$. The charge relaxation resistance is given by
\begin{equation}
R=
\frac{h}{2e^{2}}\left( \frac{\sum _{i ={\rm odd}}D_{i}^{2}}{(\sum _{i ={\rm odd}}D_{i})^{2}} +
\frac{\sum _{i ={\rm even}}D_{i}^{2}}{(\sum _{i ={\rm even}}D_{i})^{2}}  \right) \;\; .
\label{eq8}
\end{equation}
Due to the purely capacitive coupling the low frequency
admittance of the Corbino disc is 
\begin{equation}
G(\omega)=  -i\omega C_{\mu}(\omega)= -i \omega C_{\mu}(0)+ \omega ^{2}R C_{\mu}^{2}(0) \;\;.
\label{eq9}
\end{equation}
As one expects, this result is equivalent to the low frequency admittance of a mesoscopic
capacitor with $m$ channels on each side.\cite{BTP1} On the other hand, one finds for the bar
\begin{equation}
G(\omega)=  m\frac{e^{2}}{h}+i\omega C_{\mu}(\omega)= m\frac{e^{2}}{h}+i \omega C_{\mu}(0)-
\omega ^{2}R C_{\mu}^{2}(0)
\label{eq10}
\end{equation}
While the emittance of the disc behaves like a capacitance, the emittance of the bar is
inductive-like. Inductive behavior is a typical property of conductors where transmission between
different reservoirs dominates.\cite{BJPC} The crossover between the two limits of
a purely transmitting and a purely capacitive arrangement can be investigated by considering tunneling
between edge channels. In a recent work, \cite{CB2} we have shown that
the emittance of a quantum point contact without magnetic field exhibits
a change of the sign as a function of the transparency.\\ \indent 
Note that if all $D_{j}$ are of the same order of magnitude, the charge relaxation resistance
$R$ scales like $1/m$. Nevertheless, if $C_{\mu}$ scales
with $m$ (as is the case on a Hall plateau and for large $C$) one has $G(\omega)\propto m$.
At a transition between two Hall plateaus,
however, the density of states of the innermost edge channel diverges
and one expects the charge relaxation
resistance to be of the order of a single resistance quantum.  
\subsection{Four-terminal devices}
\label{Four-terminal devices}
Figure \ref{fig1}c. shows a two-terminal bar with two additional symmetric
gates each located close to an edge state. Macroscopic gates have huge
densities of states which means that $C/D_{2}= C/D_{4}=0$, where
$C=|C_{12}|=|C_{34}|$ is the geometric capacitance between a gate and
the nearest edge channel. Note that the off-diagonal elements of the
capacitance matrix are negative. We assume that the other off-diagonal capacitance elements
are much smaller, i.e. $|C_{24}|= C_{g}\ll C$ and $C_{13}=C_{14}=0$.
Now, let us ask for the current through contact $1$
in response to a voltage oscillation at gate $2$. We find to leading order of $C_{g}$
for the associated admittance 
$G_{12}|_{B}=\omega ^{2} (h/2e^{2})C_{g}C_{3} + {\cal O}(\omega ^{3})$ with
$C_{3}=1/(C^{-1}+D_{3}^{-1})$. Clearly, since edge channel $3$ emits
into reservoir $1$, and gate $2$ interacts only via gate $4$ with this edge channel,
the response is proportional to $C_{g}C_{3}$ and of second order in frequency.
On the other hand, by inverting the magnetic field, edge channel $1$ emits into
reservoir $1$. Now, the relevant capacitance turns out to be $C_{1}=1/(C^{-1}+D_{1}^{-1})$;
to leading order of $C_{g}$ we find $G_{12}|_{-B}=i\omega C_{1}+ \omega^{2}(h/e^{2})
C_{1}^{2}+{\cal O}(\omega ^{3})$. Thus while $G_{12}|_{-B}$ exhibits a capacitive response
given by $C_{1}$, the capacitive response in $G_{12}|_{B}$ is absent. Such a dramatic
magnetic field asymmetry has been verified in an experiment by Chen et al. \cite{Chen1} 
in the absence of gate $4$, i.e. for $C_{g}=0$.\\ \indent
The result (\ref{eq7}) implies that the conductance $G_{\alpha \beta}$ between
two contacts which are capacitively coupled to the sample is a symmetric function
of the magnetic field. This is in general true up to ${\cal O}(\omega)$.\cite{BJPC}
To higher order in frequency, however,
the magnetic-field symmetry of the current response can change in the presence
of voltage probes. As an example, we assume $C_{g}=0$
and that contact $3$ in Fig. \ref{fig1}c. acts as a voltage probe. Using 
$\delta I_{3}=0$, we eliminate $\delta V_{3}$ from the Eqs. (\ref{eq1}) and
obtain the reduced admittance
matrix $\tilde G_{\alpha \beta}$ associated with the contacts $1$, $2$, and $4$. We find
$\tilde G_{2 4}|_{B}=0$ and
$\tilde G_{4 2}|_{B}= \tilde G_{2 4}|_{-B}=\omega ^{2}(h/e^{2})C^{2}$. Hence, 
the impedance of a purely capacitive arrangement is in general not symmetric but
only governed by the Onsager-Casimir symmetry relations,
$G_{\alpha \beta}|_{B} =G_{\beta \alpha}|_{-B}$. This conclusion has been drawn 
by Sommerfeld et al.\cite{Som1} who found an asymmetric
capacitance of a three-terminal Corbino disc. Our theory predicts
such an asymmetric behavior in a simple way.\\ \indent
Another interesting question concerns the Hall resistance
$R_{H}=(\delta V_{2}-\delta V_{4})/\delta I_{1}$ of the sample shown 
in Fig. \ref{fig1}c. If $C_{g}$ vanishes
one expects exact quantization because the gates do nothing but measure the
edge potentials. An expansion to first order in $C_{g}$ yields the low
frequency resistance
\begin{eqnarray}
R_{H} & = & \frac{h}{e^{2}}\:(1-C_{g}(C_{1}^{-1}+C_{3}^{-1})) \nonumber \\
 & + & \frac{1}{2}\omega ^{2}\frac{h^{3}}{e^{6}}
\; \frac{C_{g}(C_{1}^{-1}+C_{3}^{-1})}
{D_{1}^{-1}D_{3}^{-1}+C^{-1}(C^{-1}+D_{1}^{-1} +D_{3}^{-1})}
+{\cal O}(\omega ^{3})
\label{eq10a}
\end{eqnarray}
In the case where the two gates are decoupled, one concludes
indeed that the resistance is quantized, $R_{H}(\omega)= h/e^{2}$.
However, due to the Coulomb coupling between the gates this result is modified
even for the dc case.
%

The integer quantum Hall effect\cite{Klit1}  
corresponds to the quantization of the Hall resistance and
the vanishing of the longitudinal resistance of the ideal
four-probe quantum Hall bar of Fig. \ref{fig1}d.
at zero frequency. Two of the contacts
serve as voltage probes, whereas the two remaining contacts are used
as source and sink for the current. With the help of
our theory the dc-results can be extended to the low-frequency case.
If the contacts $3$ and $4$ are
the voltage probes, the longitudinal resistance is defined
by $R_{L}=(\delta V_{4}-\delta V_{3})/ \delta I_{1}$.
The Hall resistance is defined by $R_{H}=
(\delta V_{2}-\delta V_{4})/\delta I_{1}$, provided the contacts $2$ and $4$
are voltage probes. Let us assume that the elements $C_{\mu , kl}$ of the 
electrochemical capacitance matrix are known.
For the specific geometry of Fig. \ref{fig1}d. we assume
$C_{\mu,12}=C_{\mu,24}=C_{\mu,34}= 0$ and calculate the low
frequency admittance. From Eqs. (\ref{eq1}), (\ref{eq6}), and
(\ref{eq7}) we find then the longitudinal resistance
\begin{equation}
R_{L} =i\omega \: \frac{h^{2}}{e^{4}} C_{\mu,13} +
\frac{\omega ^{2}}{2}\frac{h^{3}}{e^{6}}C_{\mu,13}(C_{\mu,23}-C_{\mu,14})\;\;.
\label{eq11}
\end{equation} 
The leading term of the longitudinal resistance is
determined by the Coulomb coupling between the current
circuit and the voltage circuit which are represented by edge channels
$1$ and $3$, respectively. This is even true for finite
$C_{\mu, 24}$.\cite{CB1} Moreover, for the specific symmetry
$C_{\mu,23}=C_{\mu,14}$ the dissipation vanishes also in ${\cal O}(\omega ^{2})$.
For the Hall resistance one finds 
\begin{equation}
R_{H}=\frac{h}{e^{2}} -i\omega \: \frac{h^{2}}{e^{4}} C_{\mu,13} +
\frac{\omega ^{2}}{2}\frac{h^{3}}{e^{6}}C_{\mu,13}(C_{\mu,23}+C_{\mu,14})\;\;.
\label{eq12}
\end{equation} 
We mention, that if we had chosen the symmetry\cite{CB1}
$C_{\mu ,13}=C_{\mu ,24}$, the imaginary part of the Hall resistance
would have vanished in first order of $\omega$.

\section{Conclusion}
\label{C}
We presented a theory of ac-conductance for quantized Hall
samples which takes the Coulomb interaction into account self-consistently.
It is based on a discretized model where each edge channel
is described by a homogeneous mesoscopic conductor.
The whole sample is viewed as an electric entity and the results satisfy the
important requirements of gauge invariance and current conservation.
Obviously, the theory provides a manifold of experimental tests. 
The internal dynamics enters by a single relaxation process associated
with the uniform charging or decharging of edge channels with a certain
relaxation time. A next step should be to include internal dynamics
due to spatio-temporal excitations\cite{Taly} along the edge channel within a microscopic
theory.

\subsection*{Acknowledgments}
This work was supported by the Swiss National Science Foundation under grant
Nr. 43966.


\end{document}